%% file: revtex.tex
\documentclass[%
reprint,
superscriptaddress,
amsmath,amssymb,
aps,longbibliography,
prb,
]{revtex4-2}

\usepackage{graphicx}
\usepackage{verbatim}
\usepackage[version=4]{mhchem}
\usepackage{xcolor}
\usepackage{physics}
\usepackage[normalem]{ulem}
\usepackage{pdfpages}
\usepackage{pgffor}

\usepackage[capitalise]{cleveref}
\usepackage{bm}

\usepackage{booktabs}
\usepackage{tabularx}
\usepackage{multirow}
\DeclareUnicodeCharacter{2212}{\ensuremath{-}}
\usepackage[english]{babel}
\makeatletter
\AtBeginDocument{\let\LS@rot\@undefined}
\makeatother

\begin{document}

\title{Mechanistic study of mixed lithium halides solid state electrolytes}

\input{authors}

\date{\today}

\begin{abstract}
\input{abstract}
\end{abstract}

\maketitle

\input{body-resub}

\begin{acknowledgments}
\input{acknowledgements}
\end{acknowledgments}

\input{DataAvalilability}

\bibliography{others,biblio,LXM,LPSsurface_paper}

\end{document}

%% file: authors.tex
\author{Davide Tisi}
\affiliation{Laboratory of Computational Science and Modeling, Institut des Mat\'eriaux, \'Ecole Polytechnique F\'ed\'erale de Lausanne, 1015 Lausanne, Switzerland}

\author{Sergey Pozdnyakov}
\affiliation{Laboratory of Computational Science and Modeling, Institut des Mat\'eriaux, \'Ecole Polytechnique F\'ed\'erale de Lausanne, 1015 Lausanne, Switzerland}

\author{Michele Ceriotti}
\email{michele.ceriotti@epfl.ch}
\affiliation{Laboratory of Computational Science and Modeling, Institut des Mat\'eriaux, \'Ecole Polytechnique F\'ed\'erale de Lausanne, 1015 Lausanne, Switzerland}





\definecolor{Grey}{rgb}{0.50,0.50,0.50}
\definecolor{Blue}{rgb}{0.00,0.00,1.00}
\definecolor{Red}{rgb}{1.00,0.00,0.00}
\definecolor{Green}{rgb}{0.20,0.80,0.20}
\definecolor{Magenta}{rgb}{0.60,0.00,0.60}
\definecolor{BluBondi}{rgb}{0.00,0.58,0.71}
\definecolor{Orange}{rgb}{0.95,0.46,0.17}
\definecolor{tangerine}{rgb}{0.944,0.522,0}
\definecolor{violet}{rgb}{0.5059,0.3137,0.7098}
\newcommand{\editor}[2]{%
  \expandafter\newcommand\csname #1note\endcsname[1]{%
    \textcolor{#2}{{\it (\textbf{#1:} ##1)}}}%
  \expandafter\newcommand\csname #1\endcsname[1]{%
    \textcolor{#2}{##1}}%
  \expandafter\newcommand\csname #1cancel\endcsname[1]{%
    \textcolor{#2}{\sout{##1}}}%
  \expandafter\newcommand\csname #1change\endcsname[2]{%
    \textcolor{#2}{\sout{##1} ##2}}%
  \newenvironment{#1text}{\color{#2}}{\color{black}}
}
\editor{resub}{red}
\editor{DT}{tangerine}
\editor{MC}{Green}

%% file: abstract.tex
Lithium halides with the general formula \ce{Li_xM_yX6}, where M indicates metal ions and X halide anions are very actively studied as solid-state electrolytes, because of relatively low cost, high stability and Li conductivity. 
The structure and properties of these halide-based solid electrolytes (HSE) can be tuned by alloying, e.g. using different halides and/or transition metals simultaneously. The large chemical space is difficult to sample by experiments, making simulations based on broadly applicable machine-learning interatomic potentials (MLIPs) a promising approach to elucidate structure-property relations, and facilitate the design of better-performing compositions.
Here we focus on the \ce{Li3YCl_{6x}Br_{6(1-x)}} system, for which reliable experimental data exists, and use the recently-developed PET-MAD universal MLIP to investigate the structure of the alloy, the interplay of crystalline lattice, volume and chemical composition, and its effect on Li conductivity.  
We find that the distribution of Cl and Br atoms is only weakly correlated, and that the primary effect of alloying is to modulate the lattice parameter -- although it can also trigger transition between different lattice symmetries. 
By comparing constant-volume and constant-pressure simulations, we disentangle the effect of lattice parameter and chemical composition on the conductivity, finding that the two effects compensate each other, reducing the overall dependency of conductivity on alloy composition. We also study the effect of Y-In metal substitution finding a small increase in the conductivity for the C2/m phase at 25\% In content, and an overall higher conductivity for the P$\bar{3}$m1 phase.
An extended study of the effect of metal substitution, enabled by the semi-quantitative accuracy of the universal model, shows that the effect of metal-site alloying is small, indicating that alloying can be used to optimize cost, or electrochemical stability, without major impact on bulk conductivity.

%% file: body-resub.tex
\newcommand{\pthreem}{P$\bar{3}$m1}


\section{Introduction}
All-solid-state batteries (ASSBs) are at the center of scientific investigation as the principal power source for next-generation electric vehicles (EVs) \cite{Famprikis2019,Janek2016}. Compared to conventional batteries, ASSBs swap highly flammable liquid electrolytes for much safer solid-state electrolytes (SSEs), which potentially enable shorter charging times without electrolyte polarization \cite{janekChallengesSpeedingSolidstate2023} and much higher energy density \cite{jungSolidStateLithiumBatteries2019}. 
Among the many classes of SSE, the halide-based solid electrolytes (HSE), with general formula of Li$_3$MX$_6$ (M = In, Y, Sc, or Er, etc., X = Cl or Br), have gained more and more interest  \cite{Wangscience2022, Kim2021,kwak2021, Asano2018, D4QI01306A,armand2008building, Zhang2025,cheng2025NatMat} thanks to a combination of high ionic conductivity, relatively low production cost and good air/moisture stability with respect to other classes of SSE such as sulfides and oxides. HSE have been studied for almost 30 years as a possible choice for SSE, but they became popular in 2018 when Asano and collaborators \cite{Asano2018} used different synthesis methods to obtain two compounds with high room temperature (RT) ionic conductivity: 0.051 S/m for Li$_3$YCl$_6$, and 0.17 S/m for Li$_3$YBr$_6$. This work sparked an interest in halide-based SSE and showed that the use of different halogens can significantly affect conductivity and structure. In particular, Li$_3$YCl$_6$ was found in the trigonal (\pthreem{}) phase while Li$_3$YBr$_6$ was in the monoclinic (C2/m) structure. 
Different experiments agree that \ce{Li3YBr6} has higher conductivity than \ce{Li3YCl6}, but they disagree on the effect of alloying. \citet{Maas2023} finds an increase of conductivity upon doping  \ce{Li3YCl6} with Br in the  \pthreem{} symmetry, and that \ce{Li3YBr6} has the highest conductivity at a low level of Cl doping. 
They also report a considerably higher conductivity for \ce{Li3YBr6} (0.47 S/m) than the one obtained by \citet{Asano2018}. \citet{Liu2021}, instead, finds an increase in conductivity for the 1:1 alloy \ce{Li3YBr3Cl3}, that maintains a C2/m symmetry.

Despite the experimental evidence of the importance of the halogen substitution, detailed computational studies of its effect on conductivity are lacking, mainly because most studies rely on density functional theory (DFT) to obtain an accurate description of the SSE's properties \cite{CHEN2025177167,li_computational_2022,Ballal2024,artrith_machine_2019}. Quantum mechanical approaches are quite accurate and can provide insightful evidence of relevant effects \cite{zhang_targeting_2020, smith_low-temperature_2020,lepleyStructuresLiMobilities2013,Ballal2024}, but they are burdened by a high computational cost that limits their applicability to small systems not suited to capture the disorder that arises from halogen or B-site metal substitution.   
To address this cost/accuracy tradeoff one can use machine learning interatomic potentials (MLIPs), which, once properly trained on accurate quantum mechanical data, have the accuracy of DFT at a cost only marginally higher than classical force fields \cite{Deringer2021,WANG2018178,zeng2023deepmdkit,Bartok2010, Behler2007, Smith2017, schutt2022schnetpack, Rupp2012,Butler2018,fourGenBeheler,Unke2019, Batzner_NatCommun_2022_v13_p2453,PhysRevB.104.104309}. 
Accurate MLIPs can be used to study the conduction properties of SSE \cite{miyagawa_accurate_2024,tisiThermalConductivityLi2024,Staacke2022,gigliMechanismChargeTransport2024,D4TA00721B,Pegolo2022,sendekMachineLearningModeling2022,Turk2025,D2SC01306A,Dembitskiy2025,BORAL2026102154} but they have traditionally relied on the availability of an accurate and dedicated training dataset, which can be complicated to generate and can require significant computational effort, in particular for chemically diverse compounds such as \ce{Li_xM_yX6}. In recent years, so-called universal MLIPs \cite{Batatia2023MACE-MP-0, Yang2024Mattersim, Neumann2024,mazitov2025petmaduniversalinteratomicpotential} have been constructed by training on a large dataset \cite{mazitov2025MADdataset} that spans the whole periodic table. These potentials are capable of providing (semi)quantitative results in principle, out-of-the-box, without the need for further training, or with minimal fine-tuning on a small dataset for specific applications\cite{fadillah2026molecular,Bohm2026,Benedini_2025,du2025,chorna2026comparing,PhysRevLett.134.056201,chen_interfacial_2026}.

In this work, we study the effects of alloying on the structures, phase stability and conductivity of Li$_3$YCl$_{6x}$Br$_{6(1-x)}$ using a universal potential, called PET-MAD, which is based on the Point Edge Transformer (PET) architecture~\cite{pozdnyakov2023smooth} trained on the Massive Atomic Diversity (MAD) dataset~\cite{mazitov2025MADdataset}. 
Having validated the accuracy of the universal model by fine-tuning it for the Y-containing system, we then perform a systematic study of metal alloying.

\section{Methods}

\subsection{Model architecture}\label{sec:model_architecture}

The calculations in this work are made with MLIPs based on the Point Edge Transformer (PET) architecture~\cite{pozdnyakov2023smooth}, a graph neural network (GNN) that has demonstrated state-of-the-art accuracy on diverse molecular and materials benchmarks.  At each message-passing layer $l$, PET consumes and refines feature vectors, or messages, $f_{ij}^l\in\mathbb{R}^{d_{\mathrm{PET}}}$ associated with each ordered pair of atoms $i$, $j$ within the specified cutoff distance, and where $d_{\mathrm{PET}}$ is the embedding dimension chosen as hyperparameter. These messages are updated by an arbitrarily deep transformer\cite{vaswani2017attention, bergen2021systematic}, applied independently to the neighborhood of every central atom $i$. For atom $i$ at layer $l$, the transformer (shared across central atoms $i$ and unique for each GNN layer $l$) takes the collection of incoming messages $\{f_{ji}^l\}_j$ from all neighbors $j$ as input tokens, performs a permutationally covariant sequence-to-sequence transformation, and returns the set of outgoing messages $\{f_{ij}^{l+1}\}_j$ from the central atom to all the neighbors. The target property, such as potential energy, is obtained by applying feed-forward neural networks to representations $f_{ij}^l$ and summing all the contributions over bonds and layers. The base PET model, unlike other common frameworks \cite{batzner20223,WANG2018178,zeng2023deepmdkit,musaelian2022learning,Behler2007}, operates directly on the Cartesian coordinates of the
neighbors, avoiding the need for a symmetric invariant (or equivariant) descriptor, which, in turn, leads to a favorable balance between model expressivity and computational cost. Rotational invariance is learned during training through data augmentations. Recent work from Langer
\textit{et al.}\cite{Langer_2024} showed that not-symmetrized PET can learn rotational symmetry and accurately predict observables to a very high degree of rotational equivariance.

To use a potential that could handle the large chemical diversity typical of these materials, we used the recently developed universal machine learning potential PET-MAD~\cite{mazitov2025petmaduniversalinteratomicpotential}. It is a model based on the PET architecture, trained over the Massive Atomic Diversity (MAD) dataset~\cite{mazitov2025MADdataset}, which consists of
95595 structures, containing 85 elements in total (with
atomic numbers ranging from 1 to 86, excluding Astatine). PET-MAD proved to be sufficiently accurate to allow direct application to a variety of different systems, including SSE~\cite{mazitov2025petmaduniversalinteratomicpotential}.
To validate the capability of our potential on describing halides SSE, we also fine-tuned the model on a dataset of halides SSE.

\subsection{Training set construction and validation of the ML models}
\label{sec:MLPs}

We construct the training set for the ML models in an iterative fashion.
We explored the configurational space of Li$_{3}$YBr$_{6}$ and Li$_{3}$YCl$_{6}$ using the universal potential to generate structures. Then we randomly substituted some Br atoms with Cl and vice versa to completely explore the chemical space.
To evaluate reference energetics for the specific dataset, we used the same parameters used for PET-MAD, to keep the fine-tuning internally consistent. 
The MAD dataset, as reported in Ref.~\cite{mazitov2025MADdataset}, is constructed using the Quantum Espresso v7.2 package \cite{gian+09jpcm,giannozzi_advanced_2017, Urru2020,CarmineoQE}, using the PBEsol exchange-correlation functional~\cite{PerdewPBEsol}, which has good accuracy for solids, and the standard solid-state pseudopotential library (SSSP) v1.2 (efficiency set) \cite{pran+18npjcm}, with a plane-wave cutoff of 110 Ry and a charge-density cutoff of 1320 Ry.
Electronic smearing and partial occupancies were described with a Marzari-Vanderbilt-DeVita-Payne cold smearing \cite{marz+99prl}
with a spread of 0.01 Ry. In all periodic dimensions, the first Brillouin zone was sampled with a $\Gamma$-centered grid with a resolution of 0.125 \AA$^{-1}$.  

\subsection{Green-Kubo theory}

The ionic conductivities, $\sigma$, are computed via the Green-Kubo theory of linear response~\cite{Green,Kubo}, which is a practical framework for computing transport coefficients of extended systems~\cite{gigliMechanismChargeTransport2024,Tisi2021,tisiThermalConductivityLi2024,malosso2022}. For an isotropic system of $N$ interacting particles, it reads:
\begin{equation}\label{eq:GKeq}
    \sigma=\frac{\Omega}{3k_{\mathrm{B}}T}\int_0^{\infty} \langle \mathbf{J}_q(\Gamma_t) \cdot \mathbf{J}_q(\Gamma_0) \rangle\, dt,
\end{equation}
where $\Omega$ is the cell volume, $k_{\mathrm{B}}$ is the Boltzmann constant, $T$ the temperature, and $\Gamma_t$ indicates the time evolution of a point in phase space from the initial condition $\Gamma_0$, over which the average $\langle \cdot \rangle$ is performed. $\mathbf{J}_q$ is the charge flux, which depends only on the  velocities of the atoms, $\mathbf{v}_i$, and their charges, $q_i$:
\begin{equation}\label{eq:elec-currents}
    \mathbf{J}_q=\frac{e}{\Omega}\sum_{i} q_i \mathbf{v}_i .
\end{equation}
Here, the sum runs over all the atoms, $e$ is the electron charge, and the $q_i$ are equal to the nominal oxidation numbers of the atoms.~\cite{Grasselli2019}. The GK formulation can also be reformulated in another equivalent representation, the so called Helfand–Einstein (HE) formula, which  tends to be better behaved statistically~\cite{eins05anp,helfand1960,grasselli2021,Grasselli2019}. The use of GK theory for SSEs is essential to account for the correlations 
between the charge carriers and the solid matrix, as well as among the 
carriers themselves: neglecting them, as in the more common Nernst-Einstein 
(NE) estimate, can lead to sizeable discrepancies in the computed 
conductivity~\cite{Marcolongo2017,gigliMechanismChargeTransport2024}. 
The NE approximation as the single-particle 
diffusion coefficient is statistically easier to 
converge, for this reason is widely used in \textit{ab initio} MD calculations~\cite{Geng2024,Li_2024,Baktash2020}. A comparison between the 
Green-Kubo and Nernst-Einstein results for representative systems is 
reported in Sec.~SVII of the Supplementary Material.

\subsection{Molecular dynamics and Monte Carlo simulations}

 \ce{Li3YCl_{6x}Br_{6(1-x)}} has been reported in the \pthreem{} and C2/m phases, and in this work, we investigate the dependence of ionic conductivity on material composition for both phases in their fully ordered structure \cite{Schlem2020,Asano2018}. The conductivity, $\sigma$, is obtained from MD simulations using MLIPs developed in Sec.~\ref{sec:MLPs} and 
 interfaced with LAMMPS \cite{plim95jcp,LAMMPS} through \texttt{metatrain}~\cite{METATRAIN}. Two different simulation protocols are employed, depending on whether the system contains a single halide species (\ce{Li3YBr6} or \ce{Li3YCl6}) or a mixed halide composition (\ce{Li3YCl_{6x}Br_{6(1-x)}}).
For pure halide systems, we first perform a short 20~ps NpT equilibration run, followed by a 3~ns NpT production simulation from which the conductivity is extracted. For mixed compositions (as well as alloys of the non-Li element) we begin with a 20~ps equilibration in the NpT ensemble, followed by a 200~ps NpT simulation during which 100 Monte Carlo (MC) swaps between Br and Cl and/or metal atoms are attempted every 100~fs. Fig.~S4 in the SI shows the number of accepted swaps along trajectories of \ce{Li3YClBr} in the C2/m and \pthreem{} phases at different compositions, showing that the acceptance is around 12\% throughout the simulations. 
In order to extract the conductivity, we select four snapshots spaced 50 ps apart from the NpT simulations with MC swaps; each snapshot is then equilibrated for 50 ps in the NpT ensemble, followed by a 3 ns production run for conductivity evaluation.
All simulations were performed at ambient pressure and at a temperature of 300~K, consistent with the experimental conditions we use as reference.

\section{Halogen alloying}
\subsection{Validation of the potential}
We validate the bespoke potential trained for the SSEs on its test set, and compare the results with those from the universal potential. 
The zero-shot application of the PET-MAD already shows a good accuracy, in particular a RMSE of 17.34 meV/atom for energy and 86.29 meV/\AA\, for forces, corresponding to 8\% of the force standard deviation in the test set. 
After the fine-tuning procedure, the error is reduced even further, with an RMSE of 1.38 meV/atom for energy and 39.17 meV/\AA{} for forces, corresponding to 3.6\%{} relative to the force standard deviation. Fig.~S1 of the Supplementary Information contains the corresponding parity plots. 
Throughout Secs. III B–D we report results from both models: the purpose of this systematic comparison is to assess whether the zero-shot universal potential is qualitatively reliable, which, as shown below, justifies its use for the exploration of broad chemical space, in particular, the metal-site alloying in Sec. IV, where fine-tuning a dedicated model for every composition would be impractical.

\subsection{Phase stability}
\begin{figure}[tbp]            
  \centering                    
  \includegraphics[width=\linewidth]{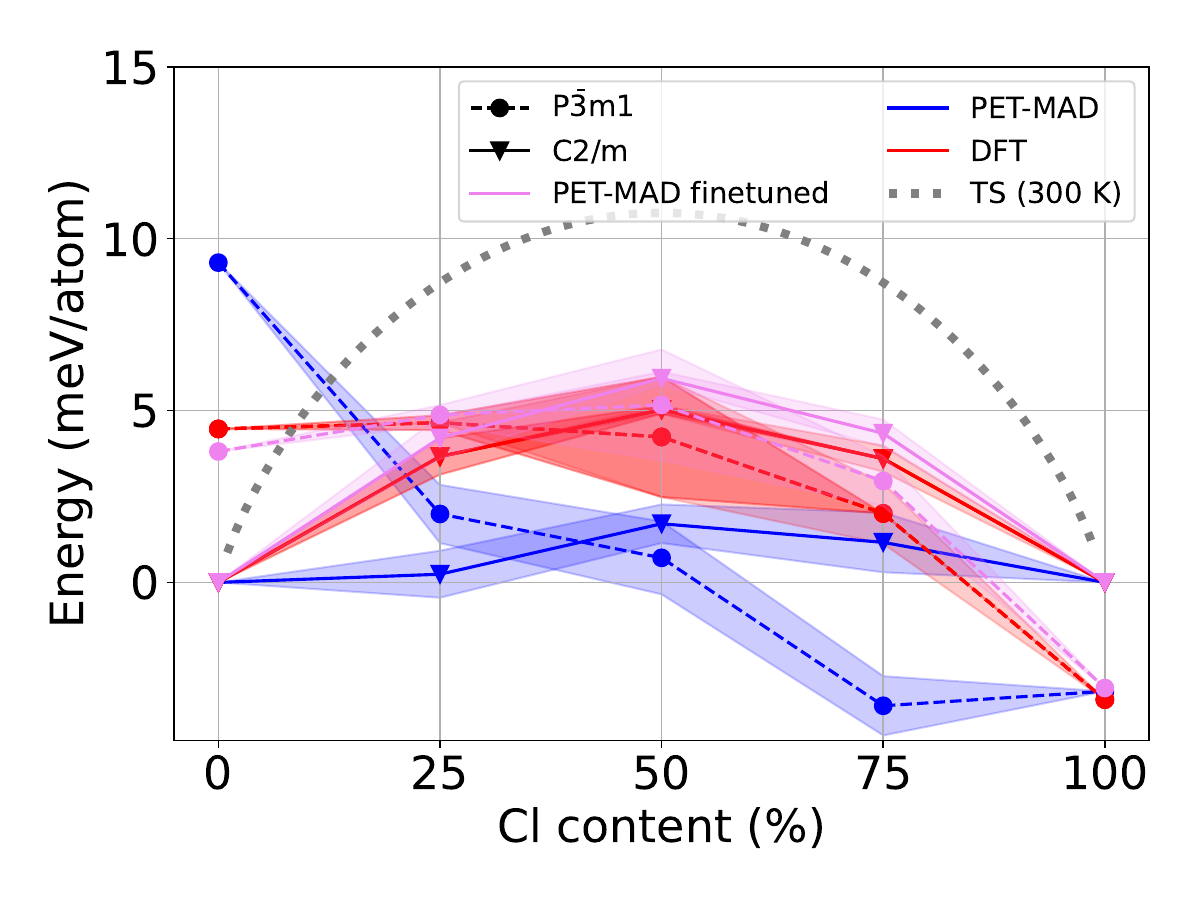}
  \caption{Formation energy for the \pthreem{} and C2/m structures as a function of the Cl content, computed with both the PET-MAD potential and its fine-tuned version. For both potentials, the baseline energies are those of the C2/m phase. For the mixed composition, the results shown are the average values of the energies of the 4 structures selected from the MC, and the shaded areas represent the semi-dispersion of the results. The dotted line represents the ideal solution entropy  $TS = -T 6  k_B  \left[ x  \ln(x) + (1 - x) \ln(1 - x) \right] $, where $x$ is the Cl fraction.  }
  \label{fig:convex_hull}        
\end{figure}

We assess the stability of the \pthreem{} and C2/m  structures upon doping. 
To this end, we quenched the snapshots we obtained from MC sampling, optimizing their geometry using both PET-MAD and the fine-tuned potentials. 
Fig.~\ref{fig:convex_hull} shows the formation energy, $\tilde{E}^{\phi}_{Cl_{6x}Br_{6(1-x)}}$, expressed relative to that of the C2/m polymorph as a function of the alloying, computed as follows:
\begin{equation}\label{eq:formationEnergy}
    \tilde{E}^{\phi}_{Cl_{6x}Br_{6(1-x)}} = E^{\phi}_{Cl_{6x}Br_{6(1-x)}} - \left[ x E^{C2/m}_{Cl_6} + (1-x)E^{C2/m}_{Br_6}\right]
\end{equation}
where $E^{\phi}_{Cl_{6x}Br_{6(1-x)}}$ is the energy of phase $\phi=P\bar{3}m1,C2/m$ and $x\in [0,1]$ is the concentration of Cl in the halide sites.
To evaluate these energies, we used the final snapshots from the 3 ns NpT production runs, then we relaxed each structure by minimizing the atomic forces, and then computed the corresponding formation energies. 
For mixed compositions, the results shown in the figure are obtained by averaging the formation energies of the 4 snapshots extracted from the MC simulation, while the shaded area is the semi-dispersion of the four results.

As shown in Fig.~\ref{fig:convex_hull}, the different snapshots have energies that differ slightly, but the relative stability of the two phases follows clear trends, that are consistent between ML models and DFT reference. 
Both the zero-shot and fine-tuned models predict the C2/m phase to be more stable in the Br-rich part of the phase diagram, in agreement with experimental observations. The crossover is predicted to occur around 1:1 halide content, which is also consistent with experimental observations. 
PET-MAD predicts a small positive mixing enthalpy for the C2/m phase, and a small negative mixing enthalpy for the  \pthreem{} phase. The fine-tuned model, instead, yields a positive mixing enthalpy for both phases. 
Comparison with single-point DFT calculations for the structure optimized with the fine-tuned model shows that -- as expected -- fine-tuning does improve the accuracy of the model.
All sets of calculations are consistent with perfect miscibility at room temperature, as even in the case of the fine-tuned model, the positive mixing enthalpy is much smaller than the ideal solution entropy (that amounts to 10.8 meV/atom at 300~K for $x=0.5$, see grey dotted line in Fig.~\ref{fig:convex_hull}).
As we shall see below, this is consistent with an analysis of the finite-temperature MC/MD, which shows that the halides are randomly distributed in the alloy.
We stress that even though fine tuning is necessary to achieve meV-level accuracy against DFT, even without this additional step PET-MAD captures correctly the qualitative behavior, namely the phase transition around 50\% {} composition, and the perfect mixing at all concentrations.

\begin{figure}[tbp]            
  \centering                    
  \includegraphics[width=\linewidth]{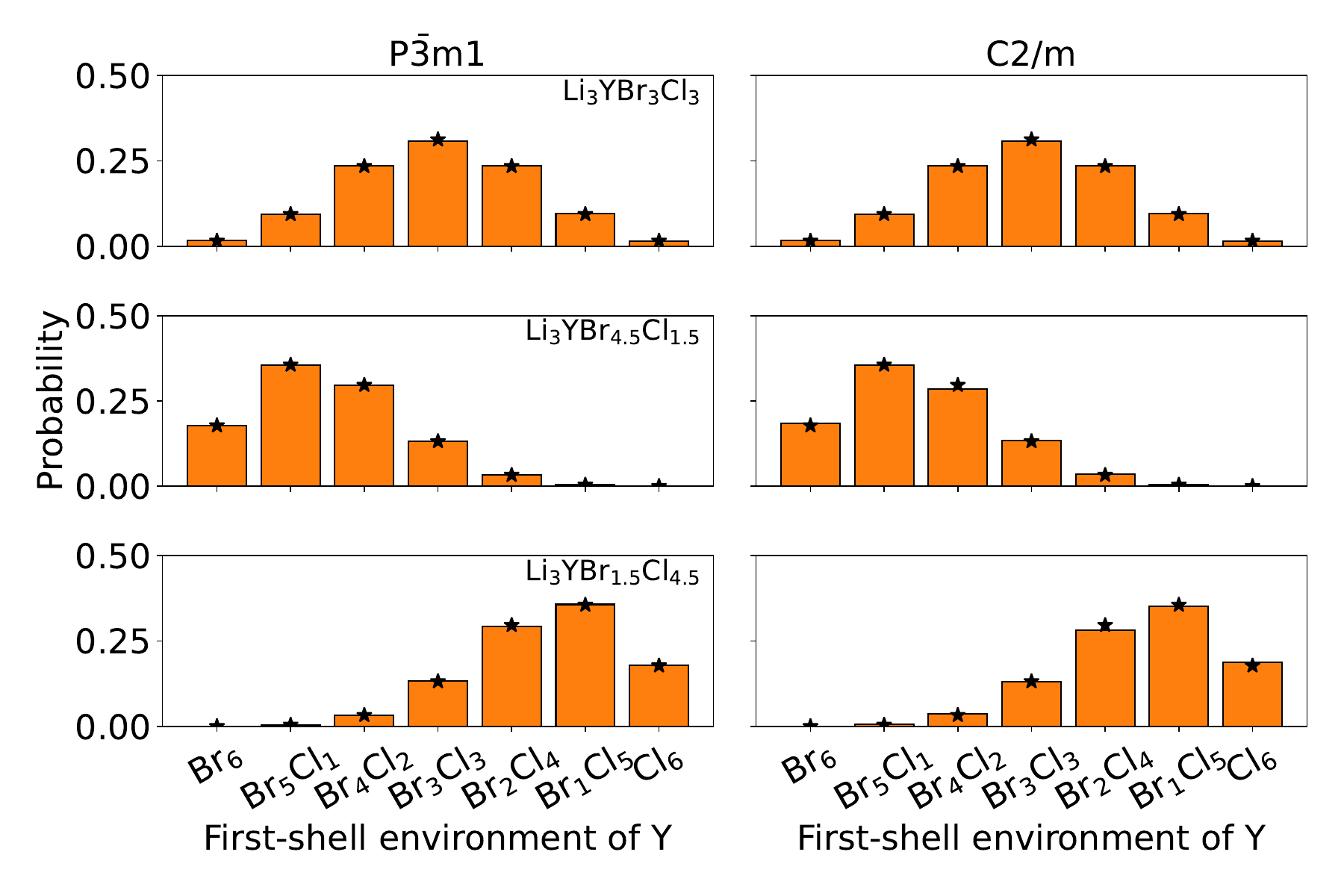}
  \caption{Distribution of octahedral compositions in the first coordination shell of Y atoms during the MC simulation for the \pthreem{} (first column) and C2/m (second column) structures. Black stars indicate the expected values from a perfect binomial distribution, with $p$ corresponding to the ratio between the number of Cl and the total number of halogen atoms in the formula units: $0.5$ for the first row, $0.25$ for the second, and $0.75$ for the third. The heights of the bars and the distributions are normalized such that the total area sums to 1. The simulations are obtained with the PET-MAD universal potential.}
  \label{fig:distribution_MC}        
\end{figure}

\begin{figure}[tbp]            
  \centering                    
  \includegraphics[width=\linewidth]{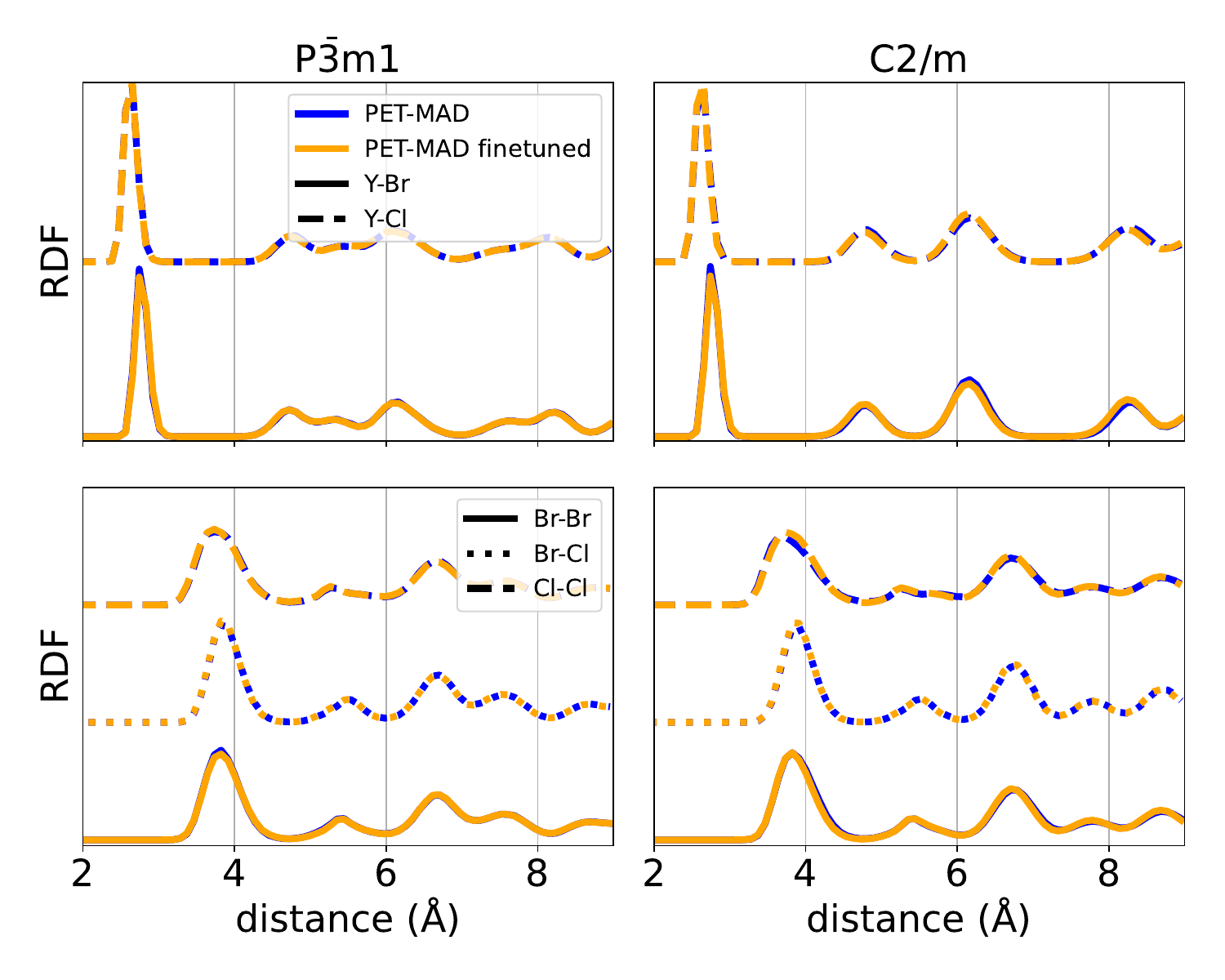}
  \caption{Radial distribution functions (RDF) of a configuration of Li$_3$YBr$_3$Cl$_3$ in the \pthreem{} phase (left) and C2/m (right) for both the PET-MAD (blue) and the fine-tuned potential (orange). The upper panels contain the RDF between the Y atoms and the halogens: Y-Br (continuous), Y-Cl (dashed). The lower panels contain the RDF between the different halogen atoms: Br-Br (continuous), Br-Cl (dotted), Cl-Cl (dashed). These are extracted from the final $50$~ps of $2$~ns long NPT simulations. In order to make the graph more readable the different components of the RDF are shifted on the y-axis by a fixed quantity.}
  \label{fig:RDF_MC}        
\end{figure}

\subsection{Structural effects of halide alloying}\label{sub:structural}

\begin{figure}[tbp]            
  \centering                    
  \includegraphics[width=\linewidth]{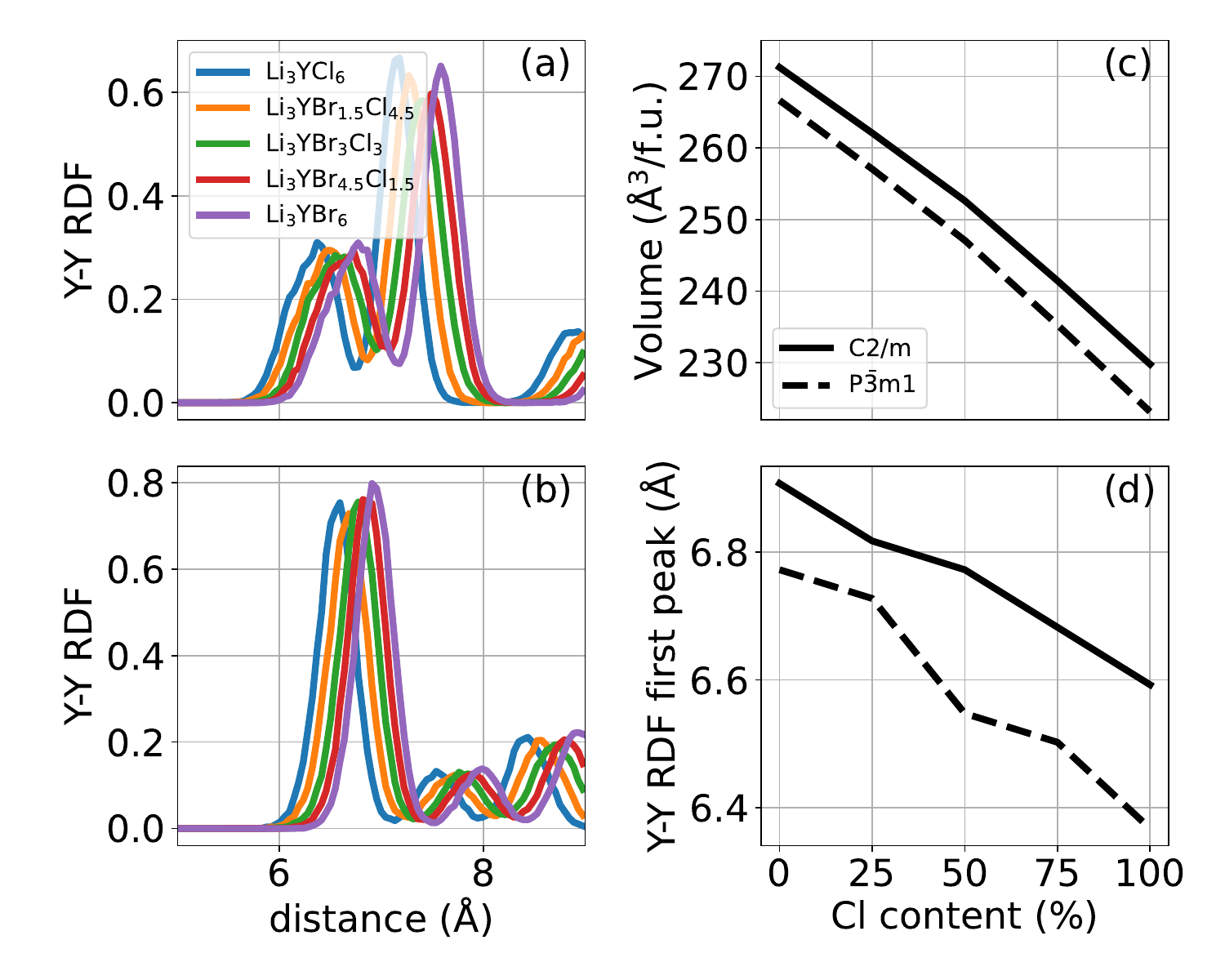}
  \caption{ Y-Y RDF for the \pthreem{} (a), C2/m (b) phases for the different chemical compositions. Behavior of the volume per formula unit (c) and the position of the first peak in the Y-Y RDF (d) as a function of the chemical composition.}
  \label{fig:RDF_chemical}        
\end{figure}

Experimental evidence shows that variations in the halide composition of SSEs affect their main structural features \cite{Liu2021}. 
However, little is known about the spatial distribution of halogen atoms within these materials, specifically, whether they tend to cluster or remain randomly dispersed \cite{D4QI01306A}. 
One way to assess the presence of short-range order in these alloys is to look at the composition of the octahedral units. In a random alloy, the number of Cl/Br anions in each octahedron should follow a binomial distribution depending on the overall composition, and any deviation would indicate the presence of correlations.
We sample structures along the trajectories of the NpT simulations, where the MC swaps are attempted for both the \pthreem{} (first column) and C2/m (second column) structures, skipping the first 10~ps for equilibration. 
The results of \cref{fig:distribution_MC} show perfect agreement between the histogram extracted from the simulations and the theoretical results, confirming that the distribution of the halides is random.

Another way to assess structural correlations is to analyze the radial distribution function (RDF).
Figure~\ref{fig:RDF_MC} presents the partial RDFs for a system of Li$_3$YBr$_3$Cl$_3$ with the \pthreem{} (left) and C2/m (right) cell with both PET-MAD and the fine-tuned model. 
Snapshots were taken every 50 ps after an initial equilibration period of 100 ps. 
The partial RDFs, shifted in Fig~\ref{fig:RDF_MC} by a constant quantity to improve readability, show that the \ce{Y-Cl} nearest-neighbor distance is slightly shorter than the \ce{Y-Br} distance, but the long-range part of the RDFs is almost identical between different halide pairs, indicating that there is no medium-range order in the alloy.
The comparison of the RDF of the 4 structures extracted from the MC simulations (see Fig. S5 in the Supplementary information) confirms that all partial halide-halide RDFs exhibit similar behaviors. This indicates that Br and Cl atoms are randomly distributed with no preference for clustering at the level of individual \ce{YX_6} units, although differences in height of the first peak suggest a slight preference for Cl atoms being located at opposing corners of the octahedron. 
The zero-shot and fine-tuned models show almost perfect agreement, underscoring the high accuracy of the PET-MAD universal model.

The radial distribution functions can also help quantify the impact of alloying on the structural parameters.
The bond length between the central metal atom in the octahedral site and the halogen atoms at its vertices remains nearly constant across different compositions. 
During molecular dynamics (MD) simulations, each structure relaxes to an average Y–Cl distance of $2.65$~\AA, and an average Y–Br distance of $2.74$~\AA. Conversely, the volume per formula unit decreases with increasing Cl content for both phases (see \cref{fig:RDF_chemical}). This contraction is also reflected in the Y–Y RDF peaks, which show that a higher Cl concentration reduces Y–Y distances, increasing the packing of the octahedra.

\subsection{Effects of alloying on the conductivity }\label{sec:alloycond}

\begin{figure}[tbp]            
  \centering                    
  \includegraphics[width=\linewidth]{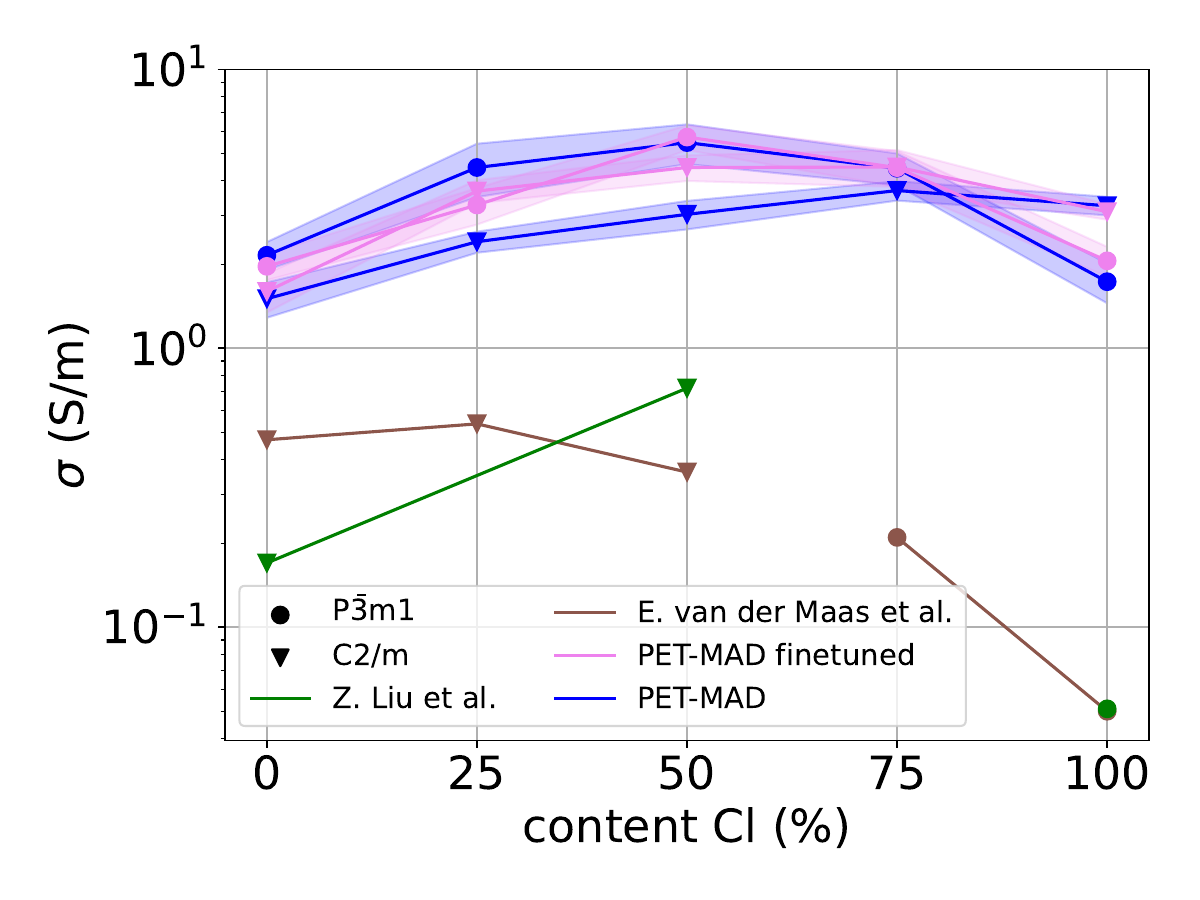}
  \caption{
  Ionic conductivity ($\sigma$) of Li$3$YBr$_{6(1-x)}$Cl$_6x$, with $x\in[0,1]$, as a function of percentage of Cl content, computed via the PET-MAD universal potential (blue) and a PET model fine-tuned (pink line) on a specific dataset. 
  The shaded area indicates the statistical uncertainty for the pure \ce{Li3YBr6} and \ce{Li3YCl6} phases; for the mixed composition, it indicates the semi-dispersion among the 4 from the different simulations selected during the MC procedure. 
  Results are compared with experimental results reported in Refs.~\cite{Liu2021,Asano2018} (green) and Ref.~\cite{Maas2023} (brown).}
  \label{fig:sigma_Cl}        
\end{figure}

We then move to consider the effect of alloying on the conductivity, performing simulations in the isothermal-isobaric ensemble, at ambient pressure and $T=300$~K.
Given that within the comparatively short time sampled by our simulations both the \pthreem{} and C2/m phases are metastable across the full range of compositions, we perform simulations in both phases and at all concentrations.
Fig.~\ref{fig:sigma_Cl} shows the result of our simulations compared to the experimental results of~\citet{Liu2021} and~\citet{Maas2023}, that (as we discussed in the introduction) are not fully consistent with each other. In Fig.~\ref{fig:sigma_Cl}, the green points for \ce{Li3YBr6} and \ce{Li3YCl6} reported in Ref.~\citenum{Liu2021} are the same conductivities obtained by \citet{Asano2018}.
The results obtained with the zero-shot PET-MAD universal potential (blue) and the fine-tuned model (pink) are in qualitative agreement with each other. 
The conductivity is overestimated by about 1 order of magnitude. The C2/m phase shows a near-monotonic increase of $\sigma$ with increasing Cl content, while the \pthreem{} has a maximum around 1:1 alloying. 
Considering the experimental (and theoretical) stability, this is consistent with the observation by~\citet{Liu2021} that the best performance can be achieved with $\approx $50\%{} doping.
Despite the large overestimation of $\sigma$, our results predict an increase in $\sigma$ also for the C2/m phase, which corroborates the experimental observations in Ref.~\citenum{Liu2021}, and not those in Ref.~\citenum{Maas2023}, that show a decreasing conductivity upon doping of \ce{Li3YBr6} with Cl. 

Another important observation from Fig.~\ref{fig:sigma_Cl} is the variability in conductivity arising from disorder in the halogen distribution. The conductivity can differ by as much as 20\%{} between the four realizations of the halogen disorder. 
Given the small energy differences among the various halogen configurations shown in Fig.~\ref{fig:convex_hull}, it is likely that all of them can form under typical synthesis conditions, such as ball milling or hot pressing \cite{Asano2018,Liu2021,Maas2023,C9EE02311A}, resulting in real materials that represent an average of these configurations.
In light of the qualitative discrepancy between experimental observations, and the large absolute shift in $\sigma$ between measurements and theory, it is clear that the most important contribution that can come from our simulations is to reveal qualitative insights -- e.g. they seem to support a maximum in $\sigma$ around 1:1 alloy composition, in agreement with~\citet{Liu2021}.
Given that in all our tests PET-MAD delivered qualitative behavior consistent with that of the fine-tuned model, we use the zero-shot universal model in the rest of our study, as it allows us to explore larger distortions and a wider portion of chemical space with no additional training cost.



\begin{figure}[tbp]            
  \centering                    
  \includegraphics[width=\linewidth]{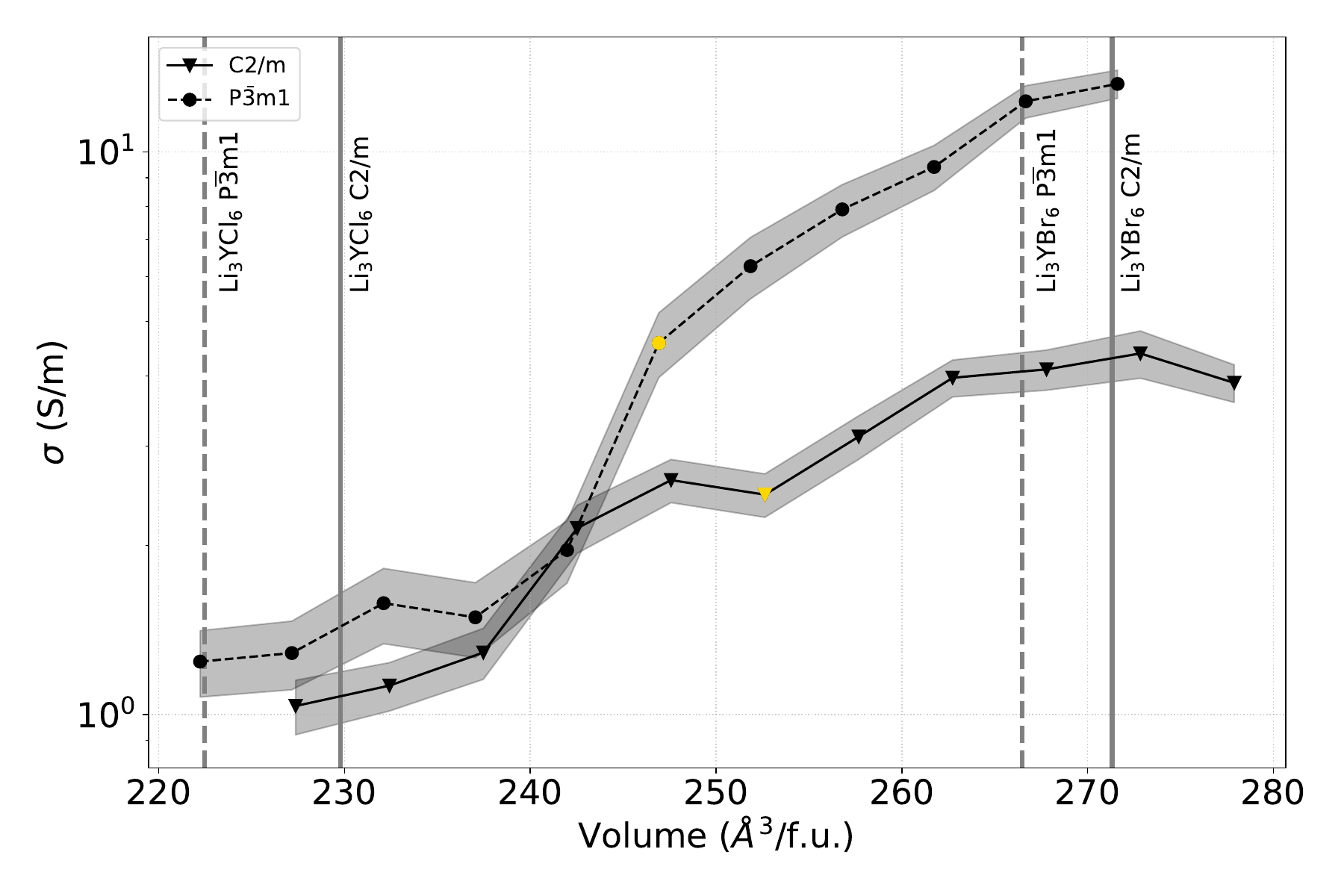}
  \caption{Ionic conductivity as a function of volume for Li$_3$YBr$_3$Cl$_3$ in the C2/m (continuous line, triangle markers) and P3m1 (dashed line, circle markers) phases. Structures were selected from the MC simulation used in \cref{fig:sigma_Cl}. For each phase, a representative snapshot with the average volume from the NPT simulation at 1 bar and 300 K was chosen. The volume was then varied by $\pm 10$ \%, and the ionic conductivity was evaluated from NVT simulations at 300 K for each volume. Yellow markers indicate the relaxed cell at 1 bar and 300 K. The vertical lines represent the average volume per formula unit, extracted from the NpT simulations, of the Li$_3$YBr$_6$ and Li$_3$YCl$_6$ for both phases.
  }
  \label{fig:sigma_vol_Y}        
\end{figure}

\subsection{Decoupling the volume and chemical composition dependence of the conductivity }

The analysis performed in Sec.~\ref{sub:structural} indicates that the main structural effect of alloying (besides modulating the stability of different phases) is to tune the metal-halide distance and the lattice parameters.
To investigate the impact of these structural changes on the conductivity of the halide-based SSE, we conduct two computational experiments. 
To study the effects of the volume, we considered one of the 50\% Cl 50\% Br structures, and modified the volume by $\pm10$\%. For each volume, we performed a NVT thermalization simulation at 300K for 200 ps followed by a 3 ns NVT simulation to compute $\sigma$. 
\cref{fig:sigma_vol_Y} shows that the ionic conductivity is highly dependent on the volume for both structures, increasing the volume by around 5\% increases $\sigma$ by a factor of 2. The potential predicts that the \pthreem{} phase has a higher conductivity at almost all volumes. 
Given that increasing Cl concentration decreases the molar volume, this effect explains why \ce{Li_3YCl_6} has a sharp increase of $\sigma$ upon alloying with \ce{Br}. 

\begin{figure}[tbp]            
  \centering                    
  \includegraphics[width=\linewidth]{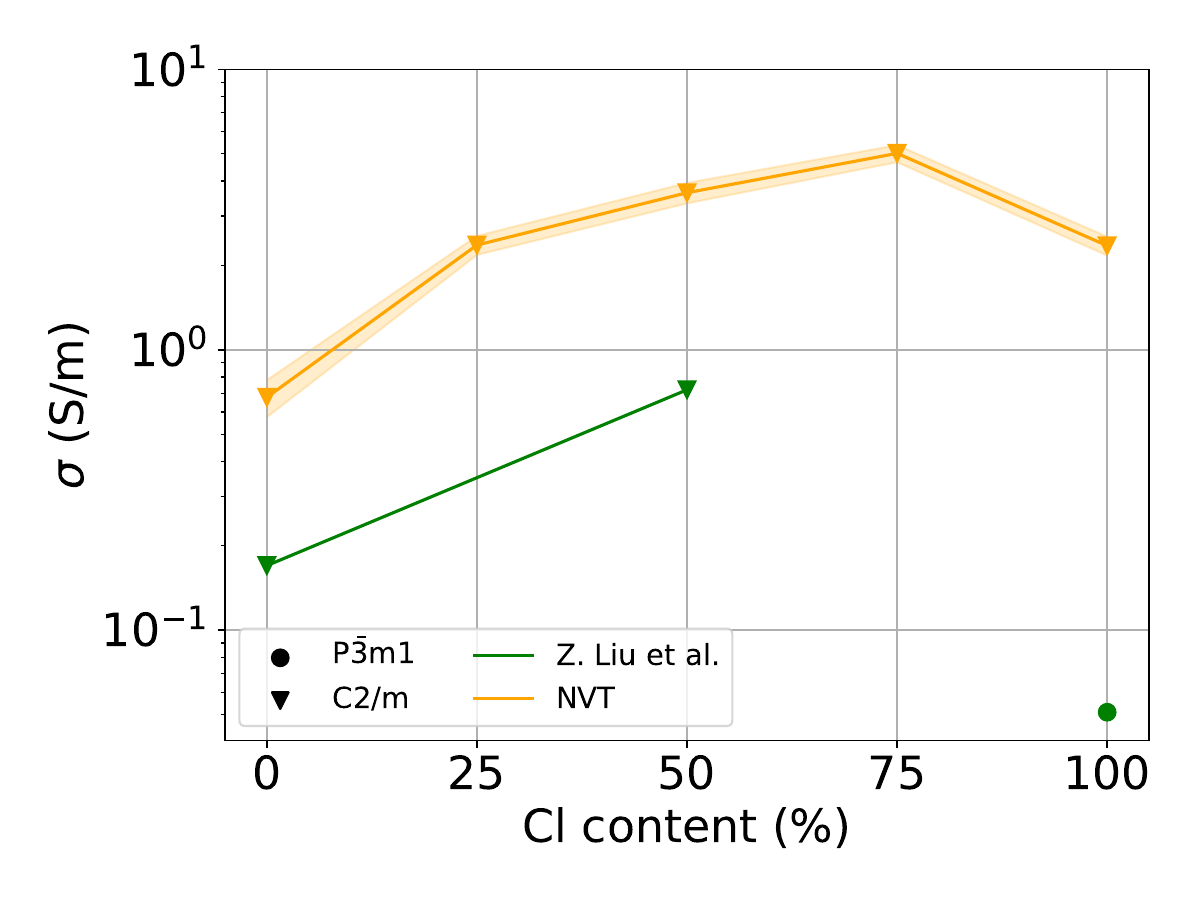}
  \caption{Ionic conductivity as a function of the percentage Cl content. Every simulation in the yellow line is done at constant volume in the NVT ensemble at the volume of the Li$_3$YBr$_3$Cl$_3$ C2/m cell of Ref.~\cite{Liu2021,Asano2018}.}
  \label{fig:sigma_NVT}        
\end{figure}

We also perform the complementary experiment, studying the impact of the different chemical compositions on the conductivity independently of the volume effects. 
We fix the volume to that of the experimental structure in Ref.~\cite{Liu2021} for the Li$_3$YCl$_3$Br$_3$ composition, then we sweep the different degrees of alloying, from full Br to full Cl substitution.
Following the same procedure described in the previous section for the mixed composition, we performed a MC swap simulation in the NVT environment, from which 4 snapshots have been selected and then used to perform separate NVT simulations (one short equilibration of 50 ps followed by a production run of 3 ns). The maximum of the conductivity is achieved for a composition around 75\% of Cl with an average conductivity that is around 8 times that obtained for the pure Br structure (\cref{fig:sigma_NVT}).

In the Br-rich part of the composition space, the two effects compensate each other, explaining why the net effect observed in the NpT calculations (Fig.~\ref{fig:sigma_Cl}) is that of increasing values of $\sigma$ up to 50-75\%{} Cl content, despite the reduction in molar volume that - alone - would reduce Li mobility. 
An interpretation of this compensation can be attempted based on the structural analysis in Sec.~\ref{sub:structural}: (1) Increasing the Cl content also leads to a contraction of the octahedral framework (as evidenced by the Y–Y peak shift in Fig.~\ref{fig:RDF_chemical}), thereby reducing Li-ion mobility; (2) However, the presence of Cl, which forms shorter Y–Cl bonds, facilitates faster Li-ion migration.
These insights provide valuable design principles for optimizing halide-based SSEs: achieving high conductivity requires balancing Cl concentration, which promotes Li diffusion, with maintaining a sufficiently large cell volume and stabilizing the \pthreem{} structure.

\section{Metal-site alloying}

\begin{figure}[tbp]            
  \centering                    
  \includegraphics[width=\linewidth]{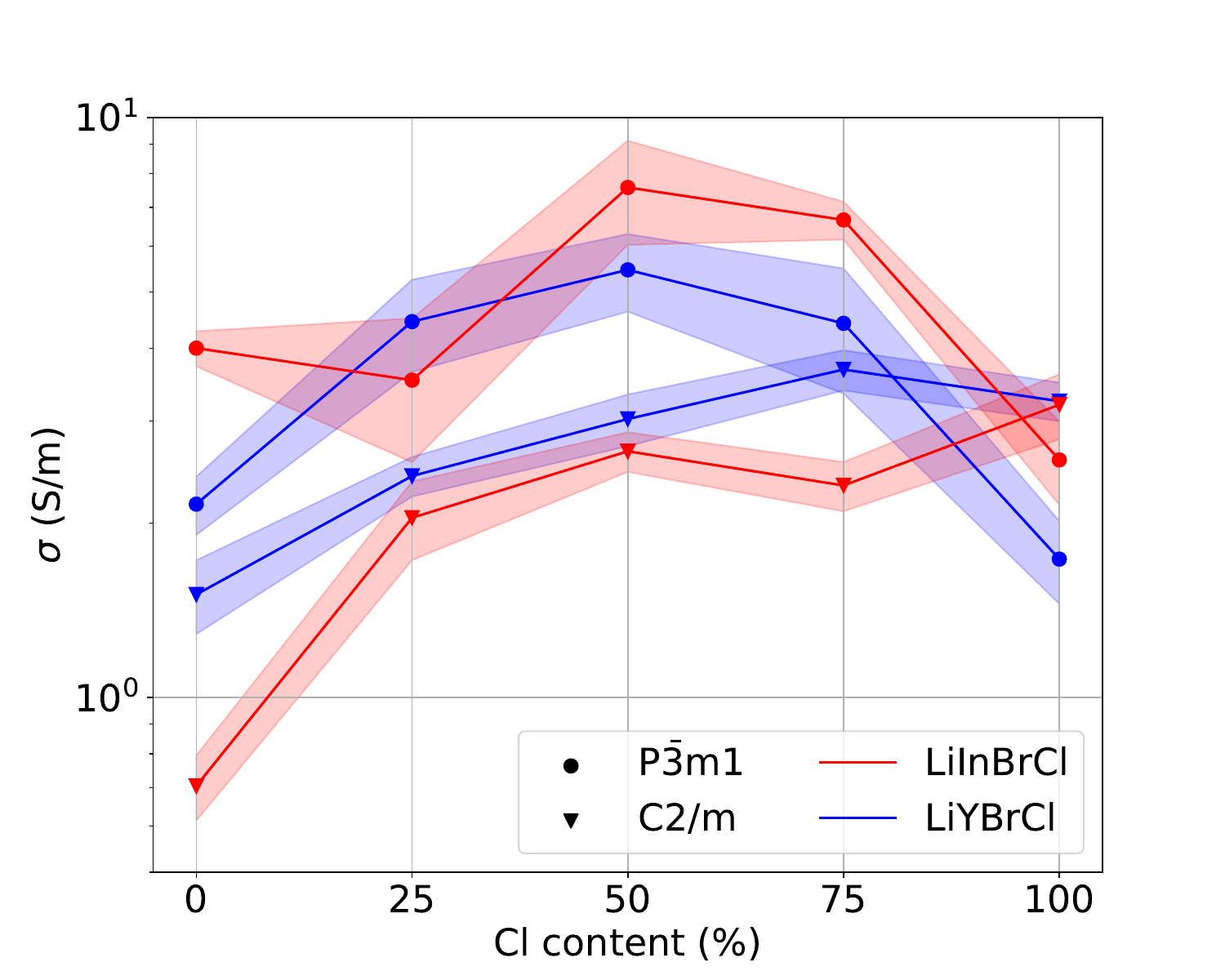}
  \caption{Ionic conductivity of \ce{Li3InBr_{6(1-x)}Cl_{6x}} (red) as a function of percentage of Cl content. Results are compared with those for \ce{Li3YBr_{6(1-x)}Cl_{6x}} (blue) shown in Fig.~\ref{fig:sigma_Cl}. All conductivities are obtained from NPT simulations, following the procedure described in the main text, using the PET-MAD universal potential.}
  \label{fig:sigma_In}        
\end{figure}

In the sections above, we have focused on the effects of halogen alloying on structure and conductivity.
On the other hand, in solid-state electrolytes the use of metal alloying~\cite{D5EB00010F,Xu2025,D5TA01706H} has also been proposed as a method to modulate the properties of the material.
The use of a universal model, such as PET-MAD, that proved qualitatively accurate against a dedicated fine-tuned model, allows us to easily study the effect of alloying the metal ion.

We begin by considering \ce{Li3InCl6}, another common halide SSE which is also found in the C2/m structure~\cite{xiong_solvent-mediated_2024}. 
Starting from the structure in Sec.~\ref{sec:alloycond},  we substitute the Y atoms with In and follow the same procedure presented in the previous sections to study the effects of halogen substitutions. 
Fig.~\ref{fig:sigma_In} shows the behavior of \ce{Li3InBr_{6(1-x)}Cl_{6x}} compared to the same behavior of \ce{Li3YBr_{6(1-x)}Cl_{6x}} from Fig.~\ref{fig:sigma_Cl}, which is qualitatively very similar. 
The \pthreem{} structure has higher conductivity, and for 1:1 halogen doping, reaches the highest value we observe. 
The C2/m phase has a lower conductivity, which increases almost monotonically in the Br$\rightarrow$Cl direction. 
As in the previous section our results can give important qualitative insights, even though they overestimate the conductivity of available experimental data, e.g. for \ce{Li3InCl6} $\sigma$ is in the $[0.6-2.4]\times 10^{-1}$ S/m range depending on processing conditions ~\cite{C9EE02311A,ROSA2026113327,li_watermediated_2019,xiong_solvent-mediated_2024}.

We then compare constant-volume and constant-pressure simulations for varying metal composition  (Fig.~S8). 
,We fix the halide composition to a 1:1 ratio, select a single realization of the halide ordering and -- for all intermediate Y-In compositions -- perform an initial MC sampling of the metal ion positions and generate four initial configurations, that are used to perform independent trajectories to estimate $\sigma$. Note that this protocol means that Fig.~S8~
can show small inconsistencies with Fig.~\ref{fig:sigma_In}, which averages over four halide configurations. 
Especially for the \pthreem{} phase, there is a substantial interplay between volume and chemical effects, with a very large increase of conductivity with In content predicted at constant volume, but a much smaller variation observed when simulating at constant pressure. 
Overall, we find a small increase of $\sigma$ for the C2/m phase, with a maximum at 25\%{} In content, while the \pthreem{} phase shows high conductivity for pure \ce{Li_3InCl_3Br_3} and \ce{Li_3YCl_3Br_3}, and lower conductivity at all intermediate concentrations. 

Having observed that the effect of halogen alloying is similar for Y and In, and that simulations must be performed at constant pressure to avoid large artifacts in the conductivity trends, we can establish a protocol to perform a more systematic exploration of the dependency of Li$^+$ ionic conductivity on B-site alloying across the binary metal pairs drawn from \{$In, Sb, Sc, Y$\}, in both the P$\bar{3}$m1 and C2/m crystal structures of Li$_3$MBr$_{3}$Cl$_{3}$.
The study of the volume dependence of the volume the chemical alloying (Fig.~S9) provides a clear signature of true solid-solution formation. In every case the cell volume interpolates essentially linearly between the two pure states, following Vegard's law \cite{DUAN20091}, showing that the two octahedral cations are mutually accommodating on a shared B-site sublattice.
Indeed, as shown in Fig.~S10, mixing enthalpies are negative, or lower than the ideal entropy of mixing, and the main thermodynamic effect of alloying is the modulation of the relative stability of the two phases.

\begin{figure}[tbp]            
  \centering                    
  \includegraphics[width=\linewidth]{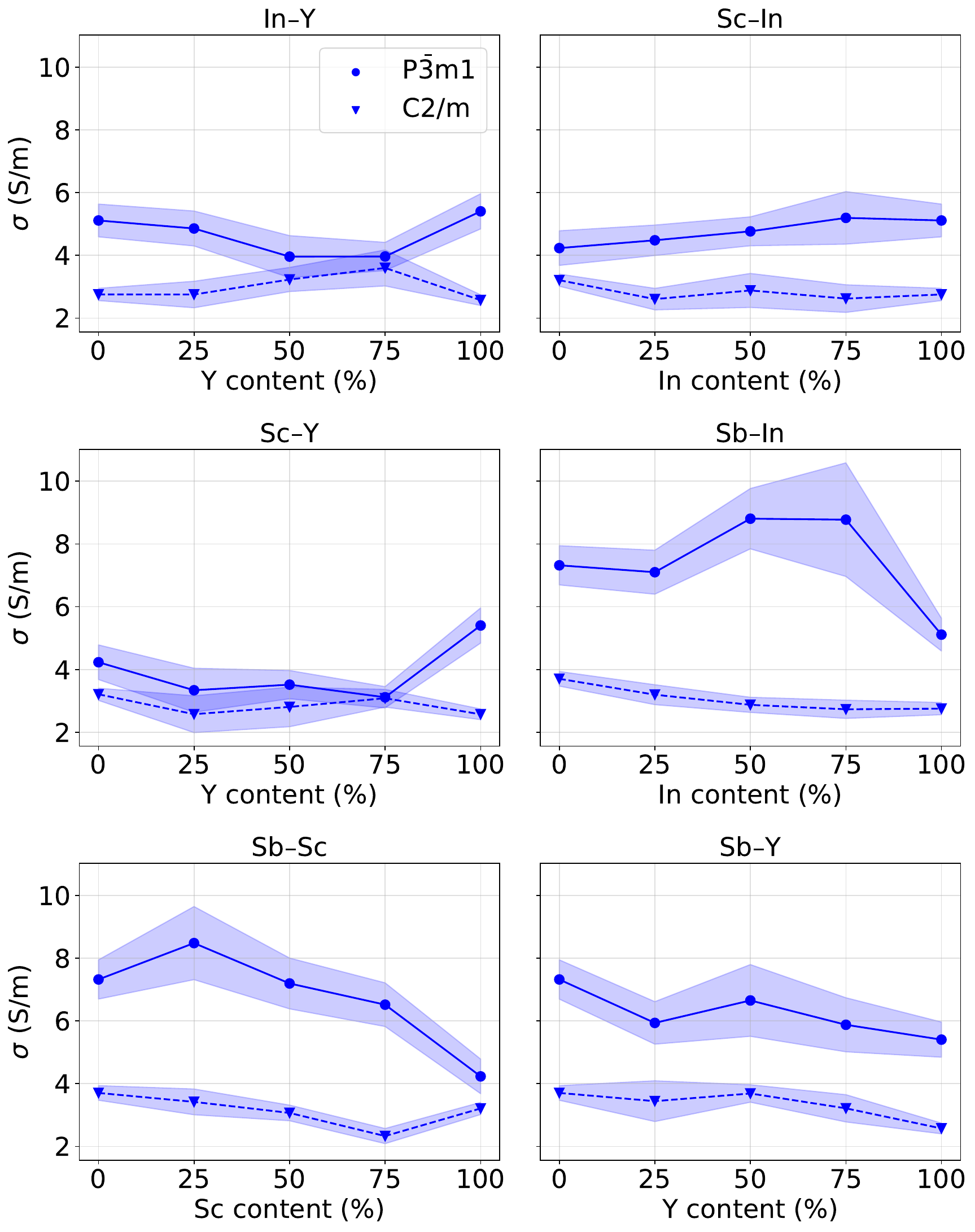}
  \caption{Ionic conductivity for the \pthreem{} and C2/m structures of the Li$_3$M1$_x$M2$_{1-x}$ (Br$_{0.5}$Cl$_{0.5}$)$_6$ as a function of the metal alloying, computed with the PET-MAD potential. Following the protocol of the paper the results at intermediate alloying are the average of 4 different metal configurations given by the Monte-Carlo swaps. }
  \label{fig:sigma_noga}        
\end{figure}

Importantly, this compositional flexibility is achieved without a substantial loss of Li$^+$ transport, as shown in Fig.~\ref{fig:sigma_noga}. Across the alloy series, the conductivity varies only moderately with B-site composition compared with the stronger compositional effects observed for the relative stability of the P$\bar{3}$m1 and C2/m phases. This indicates that metal alloying can be used as a design switch to tune phase stability, while largely preserving favourable Li-ion mobility. Moreover, because the B-site metal also might influence the chemistry, alloying provides a possible route to adjust or optimize other material properties, such as electrochemical stability and cost~\cite{Kim2021,chen2023unraveling,Xu2025,XU2024110435,kwak2021new,10.1002/aenm.202505286}, without strongly affecting bulk conductivity. 

\begin{table*}[]
    \centering
    \begin{tabular}{c c c c}
\toprule
System & Alloying at highest $\sigma$ & Stable structure & $\sigma$ (S/m) \\
\midrule
       Li$_3$(In$_{1-x}$Y$_{x}$)$_6$Br$_3$Cl$_3$ & Y$_6$ & P$\bar{3}$m1  & $5.4 \pm 0.6$ \\
       Li$_3$(Sc$_{1-x}$Y$_{x}$)$_6$Br$_3$Cl$_3$ &  Y$_6$ & P$\bar{3}$m1  & $5.4 \pm 0.6 $ \\
       Li$_3$(Sb$_{1-x}$Sc$_{x}$)$_6$Br$_3$Cl$_3$ &Sb$_6$ & C2/m  & $3.7 \pm 0.2 $ \\
       Li$_3$(Sc$_{1-x}$In$_{x}$)$_6$Br$_3$Cl$_3$ &Sc$_6$ & C2/m  & $3.2\pm0.2 $ \\
       Li$_3$(Sb$_{1-x}$In$_{x}$)$_6$Br$_3$Cl$_3$ &Sb$_6$ & C2/m  & $3.7 \pm 0.2 $ \\
       Li$_3$(Sb$_{1-x}$Y$_{x}$)$_6$Br$_3$Cl$_3$ &(Sb$_{0.5}$Y$_{0.5}$)$_6$ & P$\bar{3}$m1 & $6.6 \pm 1.1 $ \\
       \bottomrule
    \end{tabular}
\caption{Best-performing stable compositions identified from the B-site alloying survey of Li$_3$M(Br$_{0.5}$Cl$_{0.5}$)$_6$. For each binary metal pair, the table reports the alloy composition with the highest Li$^+$ ionic conductivity among the structures predicted to be stable from the relative formation energies in Fig.~S10. Conductivities are computed with the PET-MAD potential at 300 K. For intermediate alloy compositions, $\sigma$ is averaged over four metal configurations generated by Monte Carlo swaps, and the uncertainty corresponds to the semi-dispersion across these configurations.}
\label{tab:best_stable_conductivity}
\end{table*}

Overall, the P$\bar{3}$m1 phase tends to be the better Li$^+$ conductor, but also tends to have higher formation energies (see Fig. S10 in the SI). In particular only in the presence of Y alloying the P$\bar{3}$m1 is found to have lower formation energy than C2/m.  To identify the most relevant compositions from a design perspective, we therefore combine the conductivity trends in Fig.~\ref{fig:sigma_noga} with the relative phase stability reported in Fig~S10. Table~\ref{tab:best_stable_conductivity} lists, for each binary B-site pair, the composition with the highest Li$^+$ conductivity among the structures predicted to be stable. 
The best stable conductor identified in this work is the Sb–Y alloy at 50\% Y in the P$\bar{3}$m1 structure, with $\sigma = 6.6 \pm 1.1$~S/m. High stable conductivities are also obtained for the pure Y end member, which is stable in the P$\bar{3}$m1 phase reaching $\sigma = 5.4 \pm 0.6$~S/m.
Sb also reaches comparatively high conductivities in the C2/m phase, with the Sb end member having $\sigma = 3.7 \pm 0.6$~S/m.
The fact that Sb has comparatively high C2/m conductivity, and that Y stabilizes the highly-conductive P$\bar{3}$m1 phase, explains why for most of the alloying series we considered the highest conductivity is achieved for one of the pure compositions. The case of the Sb–Y series is a clear illustration of the interplay between the effect of alloying on phase stability and conductivity. The conductivity itself would reach its maximum at pure Sb for the \pthreem{} phase, but at that composition the structure is more stable in the lower-conducting C2/m polymorph. The inclusion of Y stabilizes the \pthreem{} phase, making the 50\% alloy the best performing. 
The fact that the Sb-Y series shows consistently high conductivity is practically relevant, because prices of both these elements have fluctuated widely over the past year, and so being able to modify the composition with small impact on conductivity makes it possible to optimize costs depending on market conditions. 

We have also performed simulations with Ga doping, but results show a completely different trend and are best discussed separately. 
Ga$^{3+}$ has a strong intrinsic preference for tetrahedral (GaX$_4$)~\cite{flores-gonzalez_insight_2025}
coordination rather than the octahedral (CN6) B-site geometry adopted by the
other four cations. This implies that for high Ga concentration the MX$_6$ structure breaks with the formation of GaX$_4$. This can be seen directly in the doping-induced volume
data as a structural phase transition rather than a simple alloying trend. 
As Ga content increases, $\langle V \rangle$/f.u. initially grows linearly, but beyond a threshold which depends on the specific material it jumps sharply toward anomalously large volume (see Fig. S11).
Therefore, Ga cannot be used to modulate the bulk stability and conductivity of the parent phases, but could trigger more dramatic phase transitions, and affect the macroscopic behavior by affecting the microstructure.

\section{Conclusions}

This study demonstrates the enormous potential of generally applicable machine-learning potentials for the study and design of complex functional materials, and provides mechanistic insights into the interplay between structure and composition in halide-based solid-state electrolytes. The PET-MAD potential used here achieves semi-quantitative agreement for both stability and charge transport in \ce{Li_3YCl_{6x}Br_{6(1-x)}} alloys with a more accurate model fine-tuned for this specific system, while retaining the transferability needed to explore a much broader B-site chemical space.
Both models indicate that the C2/m form is more stable for the Br-rich part of the phase diagram, with the \pthreem{} polymorph being more stable at high Cl concentrations, which is consistent with experimental observations. 
They also concur on predicting perfect miscibility of the two halides, and on the effectively random distribution of the ions, that show minimal indications of short-range ordering. 

The conductivity is substantially overestimated compared with experiments, but once again the zero-shot model and the fine-tuned model are in qualitative agreement, finding an increase in $\sigma$ with Cl content for the C2/m phase, and a maximum around the 1:1 composition for \pthreem{}.
These trends are consistent with~\citet{Liu2021}, which suggests that the decreasing conductivity seen in~\citet{Maas2023} might be indicative of effects that go beyond bulk conductivity.

An important qualitative insight we can infer is how halide substitution modulates conductivity through the interplay of different effects. 
Besides controlling the stability of different polymorphs, substitution of Br with Cl leads to a decrease in volume of the structure, which in turn reduces $\sigma$. 
This effect is, however, compensated by the contraction of the \ce{YX_6} octahedra, which leaves more space for Li diffusion despite the contraction of the lattice. 
There is also significant variability in the values of $\sigma$ depending on the sampling of the halide disorder, which makes it even more crucial to rely on MLIPs to be able to simulate large supercells and to average several configurations.

A similar interplay between molar volume and local structure is seen when considering homovalent substitution of the metal ion. The trends of $\sigma$ for Li$_3$InCl$_{6x}$Br$_{6(1-x)}$ are analogous to those seen in the Y compound, while the mixed In/Y compounds show a large composition dependence of $\sigma$ at constant volume, but much weaker variability when simulating at the equilibrium density for each composition. Extending the analysis to a more diverse collection of binary B-site alloys shows that metal substitution generally has a comparatively moderate effect on Li$^+$ conductivity, while it can substantially affect the relative stability of the P$\bar{3}$m1 and C2/m polymorphs. In the Sb–Y series, for instance, adding Y stabilizes the \pthreem{} polymorph, so that the 50\% composition becomes the best-performing stable conductor even though $\sigma$ itself is highest for pure Sb. This suggests that B-site alloying can be used to tune phase stability, electrochemical stability, redox chemistry, and materials cost, without necessarily compromising bulk Li-ion transport. Although there is a significant quantitative discrepancy between simulations and experiments, which can likely be attributed to a combination of errors in the reference DFT method and limitations in estimating conductivity for an ideal crystalline bulk, our simulations suggest several design principles. 
First, the 1:1 halide composition seems to yield high values of $\sigma$ across different polymorphs and metal compositions. Second, the molar volume and the metal-halide distances are important structural parameters that can be determined in the pure compounds and used as guides to test new alloys. Third, the weak-to-moderate sensitivity of conductivity to many B-site substitutions indicates that metal alloying is a viable strategy to optimise properties other than transport, such as polymorph stability and electrochemical compatibility, while retaining favourable Li$^+$ mobility. 
More broadly, this study highlights the value of universal machine-learning potentials: even when their quantitative accuracy is limited, their qualitatively correct trends can guide materials design across chemical spaces that would be difficult to explore exhaustively with dedicated models, not to mention computationally prohibitive first-principles simulations.

%% file: acknowledgements.tex
We thank A. Mazitov and F. Bigi for helpful discussions about the use of PET-MAD and its fine-tuning strategies. DT thanks 
Manuel Dillenz for insightful comments on a first draft of the manuscript.
DT and MC acknowledge support from a Sinergia grant of the Swiss National Science Foundation (grant ID CRSII5\_202296).
MC acknowledges support from the European Research Council (ERC) under the research and innovation program (Grant Agreement No. 101001890-FIAMMA) and the NCCR MARVEL, funded by the Swiss National Science Foundation (SNSF, grant number 205602).
This work was supported by grants from the Swiss National Supercomputing Centre (CSCS) under the projects s1243, s1219, lp26, and lp95.

%% file: DataAvalilability.tex
\section*{Data Availability}
All the data to reproduce the figures will be available on Material Cloud~\cite{TalirzMatCloud} upon publication.